\documentclass[fleqn,usenatbib]{mnras}

\usepackage{newtxtext,newtxmath}

\usepackage[T1]{fontenc}
\usepackage{ae,aecompl}

\usepackage{graphicx}	
\usepackage{amsmath}	
\usepackage{float} 
\usepackage{booktabs}
\usepackage{multirow}
\usepackage{subcaption}
\usepackage{soul}
\usepackage{color}
\usepackage{lineno}

\newcommand{\kms}{km\,s${}^{-1}$ }

\title[Young Vertical Phase]{Vertical Kinematics of the Young Galactic Clusters}

\author[Alfaro, Sánchez Gil, and Elmegreen]{Emilio J. Alfaro
$^{1}$, M Carmen Sánchez Gil $^{2}$ \thanks{E-mail: mcarmen.sanchez@uca.es}, and Bruce Elmegreen$^{3}$
\\
$^{1}$Instituto de Astrof\'isica de Andaluc\'ia, (CSIC), Glorieta de la Astronom\'ia, S/N, Granada, 18008, Spain\\
$^{2}$Universidad de Cádiz, Avenida de la Universidad N4, Jerez de la Frontera, 11406, Cádiz,  Spain\\
$^{3}$Katonah, NY 10536, USA\\
}

\date{Accepted December 2024. Received 30/09/24; in original form 30/09/24}

\pubyear{2024}

\begin{document}
\label{firstpage}
\pagerange{\pageref{firstpage}--\pageref{lastpage}}
\maketitle
%
\begin{abstract}
The young disc vertical phase is paramount in our understanding of Galaxy evolution. Analysing the vertical kinematics at different galactic regions provides important information about the space-time variations of the Galactic potential. The vertical phase snail-shell structure found after Gaia DR2 release encompasses a wide range of ages. 
However, the structure of the $V_Z\, vs \, Z$ diagram appears linear when the analysis is limited to studying objects younger than 30 Ma. 
Based on the vertical velocity and height-over-disc maps obtained for a sample of young open clusters, this method also allows the matter density in the Solar neighbourhood to be estimated using a completely different approach than previously found in the literature. 
We use two different catalogues of star clusters to confirm the previous result and study new age ranges. The linear pattern between $V_Z$ and $Z$  shows different slopes, $\partial V_Z/\partial Z$, for various age groups. The results fit a simple model (harmonic oscillator) of in-plane decoupled vertical dynamics up to a certain age limit, corresponding to $\sim$ 30 Ma. 
This work also analyses the relationship between the local volumetric density of matter ($\rho_0$) and the disc vertical kinematics for different age ranges, all below 50 Ma. The best estimates of the effective volumetric mass density in the Solar neighbourhood, 0.09-0.15 M$_\odot$ pc${}^{-3}$, agree with those given by other authors, assessing the reliability of the proposed dynamical model. These values are a minorant of the actual matter density in the region. 
\end{abstract}



\section{Introduction}\label{intro}

\par The Gaia data \citep{GaiaDR12016, GaiaDR2_2018, Gaia_EDR3_21, Gaia_DR3_23},  have provided surprising results in many branches of astrophysics. One of the most conspicuous was detecting a spiral pattern in the vertical phase space ($V_Z,\, Z$) for disc stars in the Solar neighbourhood \citep{Antoja18}. Subsequent studies have confirmed this result by extending it to various age, Galactocentric radius, and metallicity ranges \citep{Bland-Hawthorn19, Li2020, Li2021, Antoja23}. In all the cited cases, the youngest age of the analysed sample exceeded 500 Ma, and the temporal interval was over one Ga. Several explanations have been given for the generation of this conspicuous pattern, which can be divided into two main scenarios: a) the combination of a transient and highly energetic event with an asymmetric Galactic potential \citep{Antoja18, Bland2021, Tepper2022}, and b) dynamical resonances associated, in this case, with the region close to the corotation \citep{Tatiana2018, Tatiana2019, Lepine22}. In both cases, it is the dynamical effect of an asymmetric Galactic potential, with different boundary conditions, acting over a wide time span. 
V
Recently, \cite{Alfaro22} determined the 4-dimensional (X, Y, Z, V$_Z$) phase subspace structure of the young Galactic disc defined by a sample of Galactic clusters with ages less than 30 Ma. The sample (G-YOC) was obtained from these objects' position and velocity catalogues compiled from Gaia data \citep{soubiran2018,Cantat-Gaudin2020,Tarricq2021}. The vertical phase is shown by the young clusters \citep[see Fig. 12 in][]{Alfaro22} presents a pattern quite different from that of the old stars in the Galactic disc discovered by \cite{Antoja18}. The vertical phase defined by G-YOC presents a very high linear correlation between V$_Z$ and Z. The samples used in both studies are also very different in two fundamental aspects:  a) the disc stars are, on average, old objects with ages exceeding the Ga, and b)  the first detection of the phase space spiral structure \citep{Antoja18} was restricted to a cylinder with a circumference of 200 pc radius centred on the Sun, while the young clusters are distributed in a square of 7 kpc side centred on the Sun. Further studies have identified the presence of these spirals over a wide range of Galactocentric radii, although the orientation of the main axis seems to vary with the Galactic radius; a kind of apsidal motion  \citep[see][and references therein]{Antoja23}.

In this work, we want to confirm the results obtained in \cite{Alfaro22} with other samples of young clusters and propose a scenario for generating the observed linear pattern. This simple model allows us to obtain a minorant of the volumetric mass density in the Solar neighbourhood ($\rho_0$), representing a new approach to this problem. The paper is divided into five sections, the first of which is this introduction. {Section 2 presents the new astrometric data that form the initial sample of our analysis, as well as the methodology used to obtain the Kriging maps. In section 3, we describe the dynamical model and compare the properties of the observed vertical phase with those derived from the vertical acceleration model.}  Section 4 presents the proposed dynamical model and estimates the value of the effective volumetric mass density, while, finally,  section 5 discusses the results and summarises the conclusions of the work.  

\section{Young open cluster data and the observational vertical phase diagrams}\label{sec:data} 

In addition to the G-YOC sample defined in \cite{Alfaro22}, we include here a new sample obtained from the catalogue of open clusters of \cite{Dias2021}, which we call D-YOC.

\cite{Dias2021} lists a set of fundamental parameters for 1743 open clusters in our Galaxy. Information about distance, proper motions, and age has been derived from the \textit{Gaia}-DR2 data in a homogeneous way. However, the selection of cluster members has been compiled from the literature, which may include some inhomogeneities that may result in a higher internal imprecision of the derived parameters. They also include the mean radial velocity for 831 clusters, 198 of which were new additions at the time of publication. The radial velocities, together with the astrometric data from the catalogue, yield a reliable set of space velocities to determine the V$_Z$(X, Y) map of the Galactic disc in the Solar neighbourhood. This fact has led us to select the catalogue of \cite{Dias2021} to extend the study of the phase diagram of the young vertical disc.  

We take the Galactic plane as the main reference plane, and both the Cartesian space and velocity coordinates are referenced to the Sun. That is $\Vec{r}_{\odot}\equiv (X\odot,Y\odot,Z\odot)\, =(0,0,0)$ pc and $\Vec{V}_{\odot} \equiv(V_{\odot X},V_{\odot Y}, V_{\odot Z})=(0,0,0)$ \kms. The axes are positive in the usual directions: $X$ towards the Galactic centre, $Y$ in the direction of Galactic rotation, and $Z$ towards the Galactic North pole — the same for velocities. 

The selection criteria are also similar to those that defined the G-YOC sample: Clusters within a radius $ r \leq 3.5 $ kpc around the Sun, and age $ \leq 10^{7.7} $ years. Here, we extend the age upper limit by 0.2 dex with respect to G-YOC. 

With these selection criteria, we obtained a sample of 485 clusters with Galactic Cartesian coordinates and 135 with the three spatial velocities. The interest in including new cluster catalogues, constructed with different but robust methodologies, is based on the need to verify that the linear pattern in the V$_Z$, Z) diagram is an intrinsic property of the young Galactic cluster population and not an artefact resulting from a sample bias. 

On the other hand, as we will see below, the maps obtained for different age subsets will allow us to establish a more confident lower bound of $\rho_0$.

\begin{figure*}
\centering
\includegraphics[width=\textwidth, angle=0]{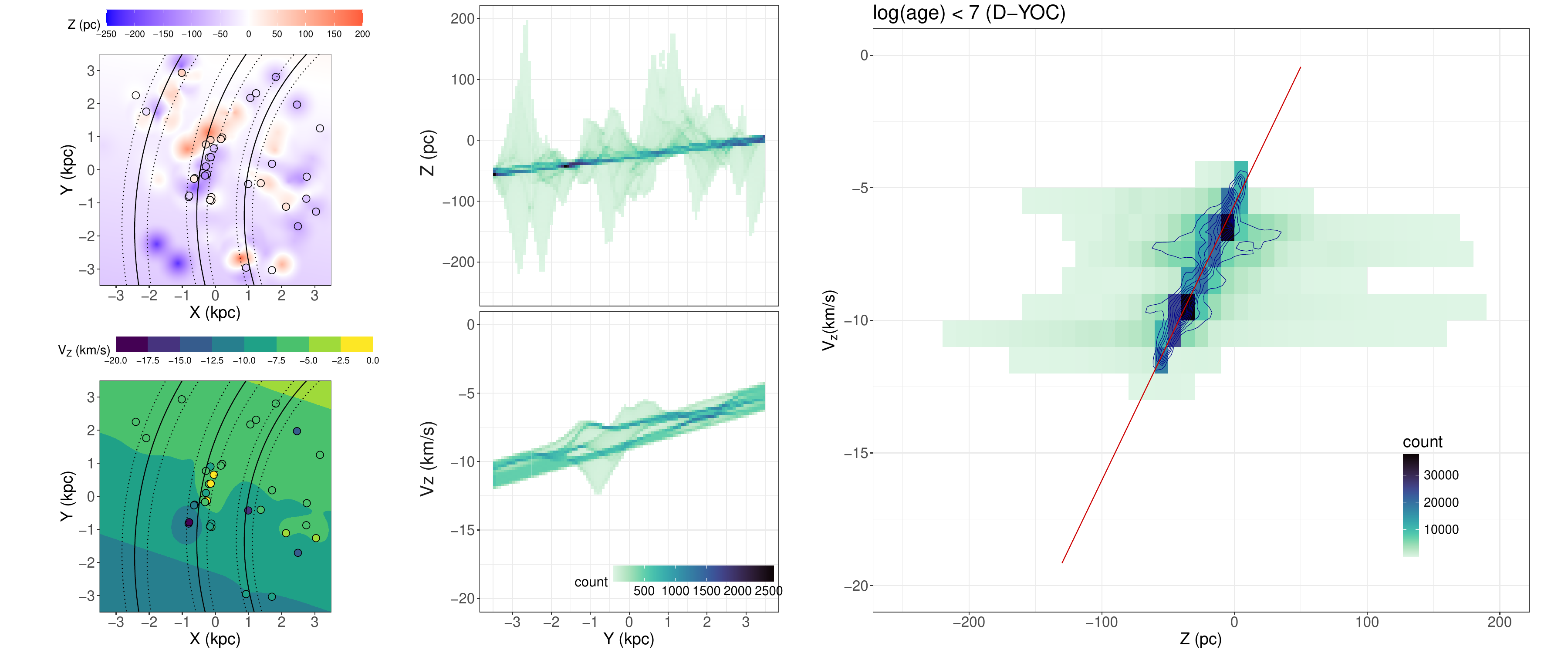}
\includegraphics[width=\textwidth, angle=0]{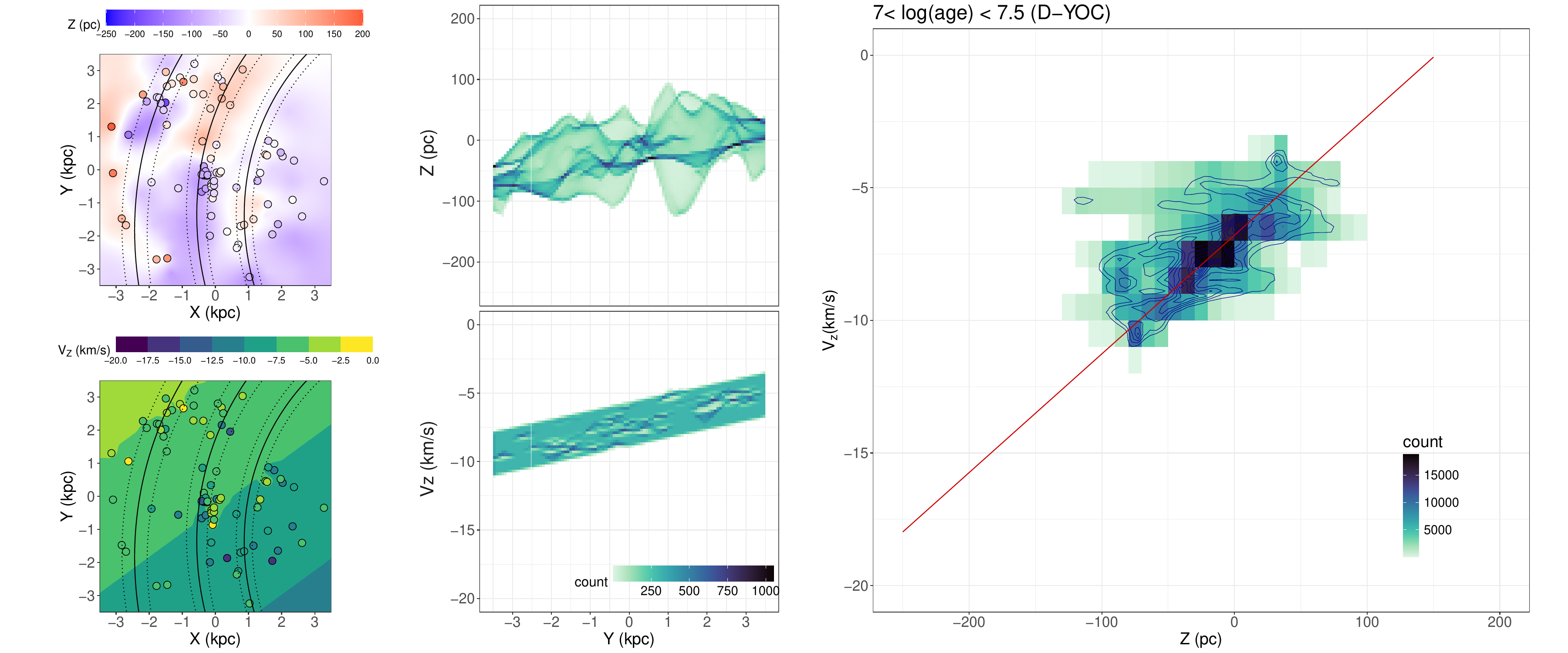}
\includegraphics[width=\textwidth, angle=0]{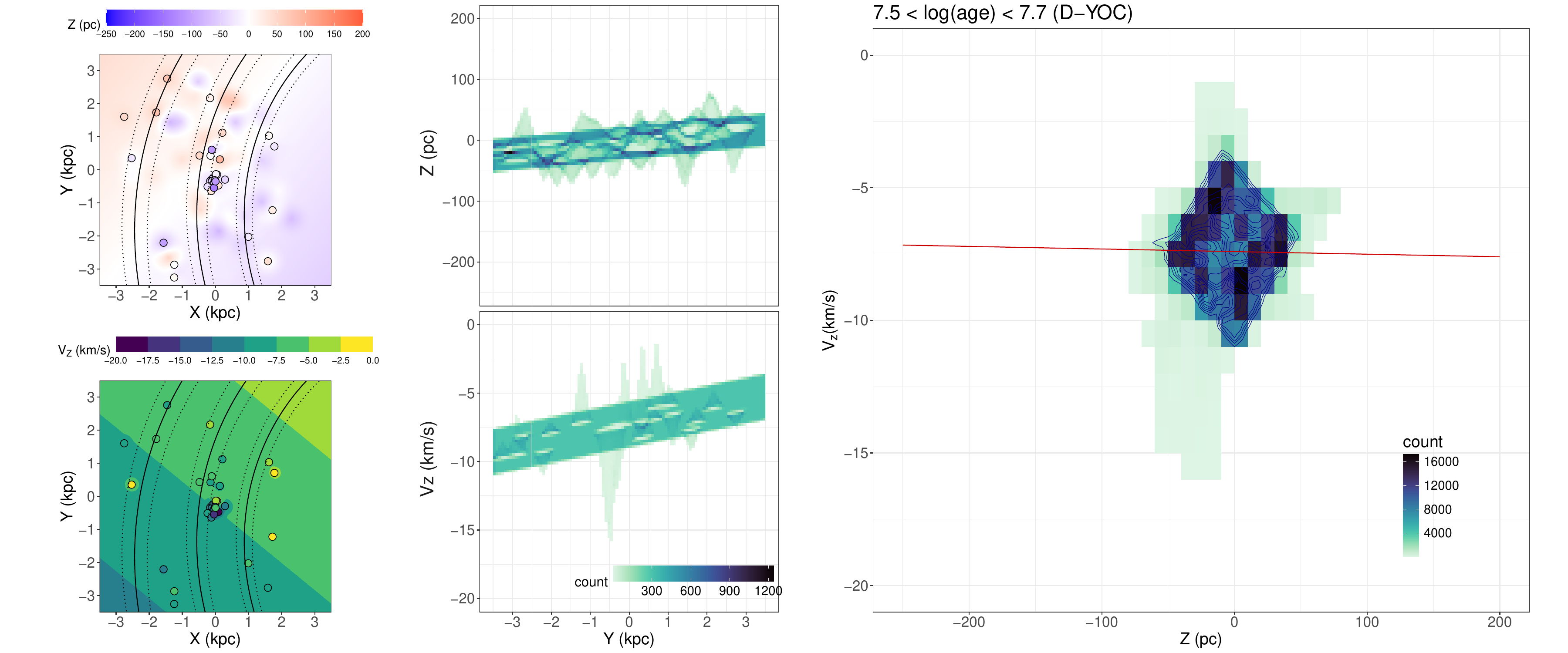}
\caption{{\it Left panels: } ${Z}(X,Y)$ and ${V}_Z(X,Y)$ Kriged estimated maps for the D-YOC sample \citet{Dias2021}, divided by age into four sub-samples, and G-YOC sample \citet{Alfaro22}. {On top of these maps, the filled points correspond with the locations and values of the original sample data}. 
Colour maps are set to a common scale to make straightforward comparisons. $V_Z(X,Y)$ were binned into a discrete scale due to their small range of values, enhancing the velocity gradients found. The three black lines correspond to, from left to right, Perseus, Local, and Sagittarius spiral arms \citep{CG2021}, with fixed arms' width based on \citet{2014R}. The Sun is at $(0,0)$.} 
\label{fig:Kriging1a}
\end{figure*}
\begin{figure*}
\includegraphics[width=\textwidth, angle=0]{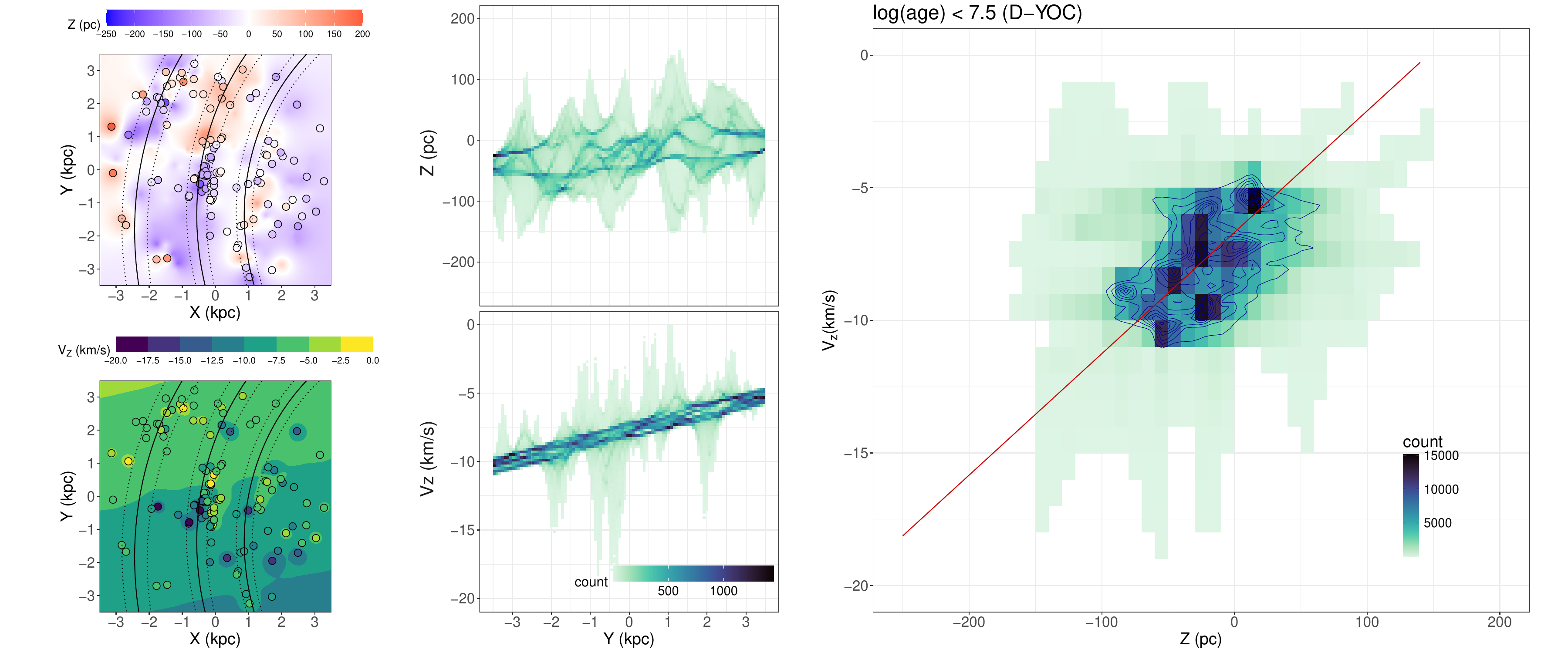}
\includegraphics[width=\textwidth, angle=0]{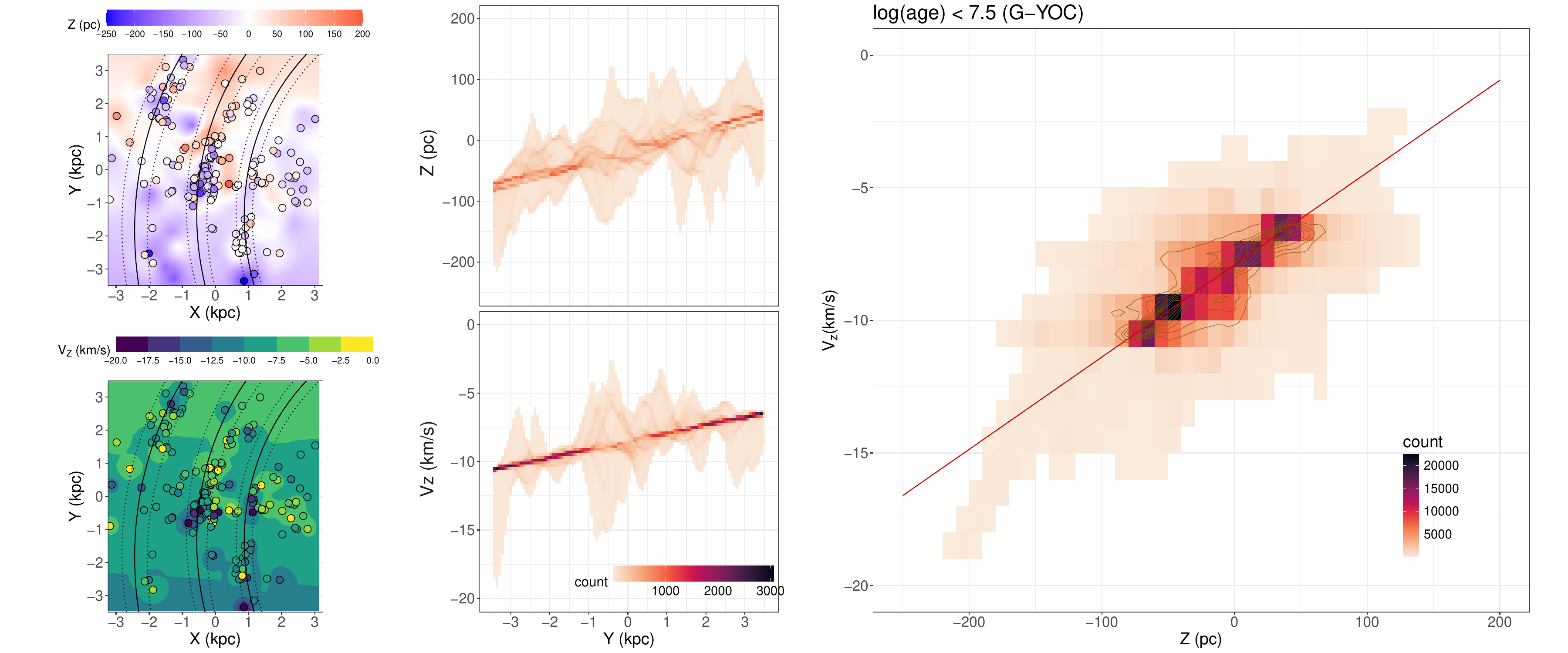}
\contcaption{
{\it. Middle and right panels: } 
The Kriged estimated ${V}_Z$ and ${Z}$ versus $Y$, and ${V}_Z$ versus ${Z}$ diagrams, respectively, for the (D-YOC) samples, represented as { heatmaps of 2D bin (of $10$ pc $\times \  1$ \kms) counts, in blue-green colour. The higher-density contour lines (in blue) are also plotted at the right panels}. Analogously for the (G-YOC) sample, in brown-orange.  
Despite the apparent wide outspread, the lighter colours correspond to values {always} below the $\mathbf{10}$ percentile of the points density distribution. The variable ranges are the same for all the plots, so we can easily compare the different age sub-samples. 
}
\label{fig:Kriging1b}
\end{figure*}

\begin{figure*}
\centering
    \begin{subfigure}{0.9\textwidth}
    \includegraphics[width=\textwidth, angle=0]{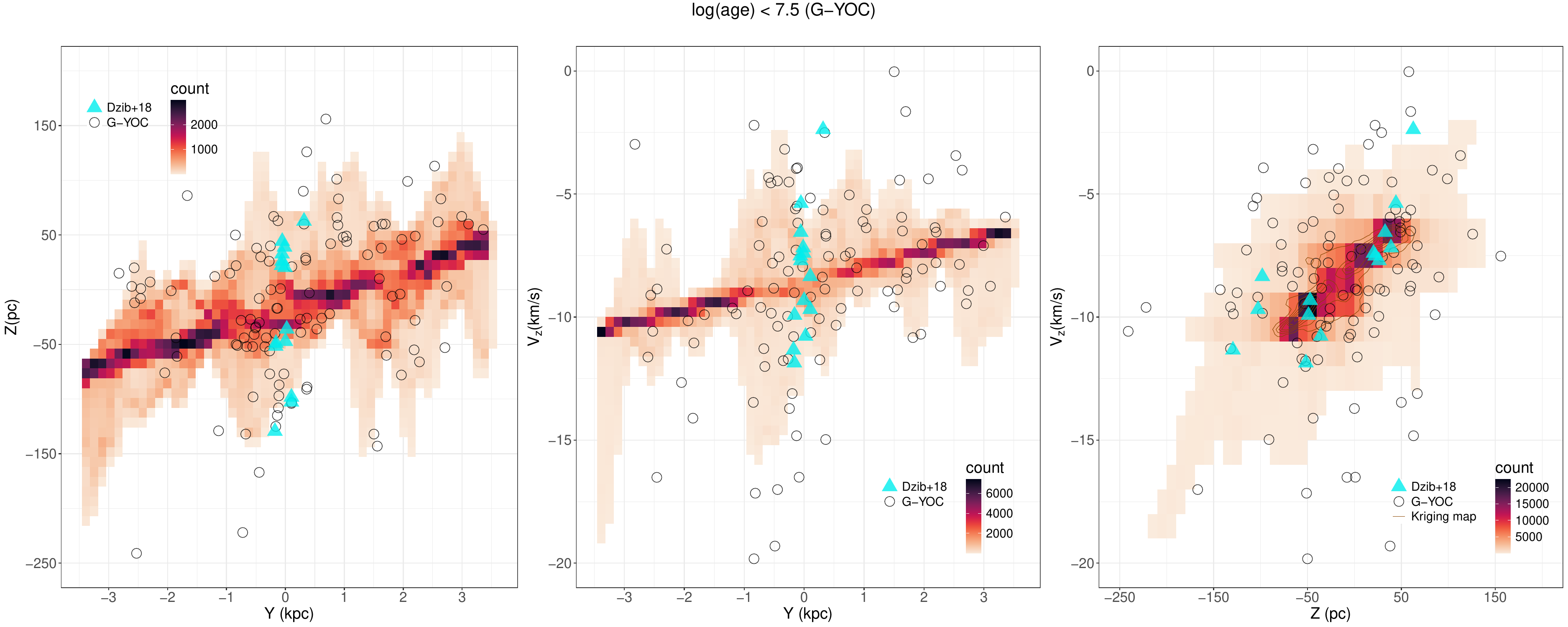}
    \caption{{ From left to right Z-Y,V$_Z$-Y, and V$_Z$-Z  diagrams were derived from the Kriging maps for the open cluster (OC) sub-sample with $log(age) < 7.5$ \citep{Alfaro22}. Black circles indicate the original G-YOC sample points, validating the agreement between the data distribution and the Kriging model. Star-forming regions near the Sun  \citep{Dzib2018} are also overplotted superimposed as cyan-filled triangles.}}
    \label{fig:Kriging2a}
    \end{subfigure}
    \hfill
    \begin{subfigure}{0.9\textwidth}
    \includegraphics[width=\textwidth, angle=0]{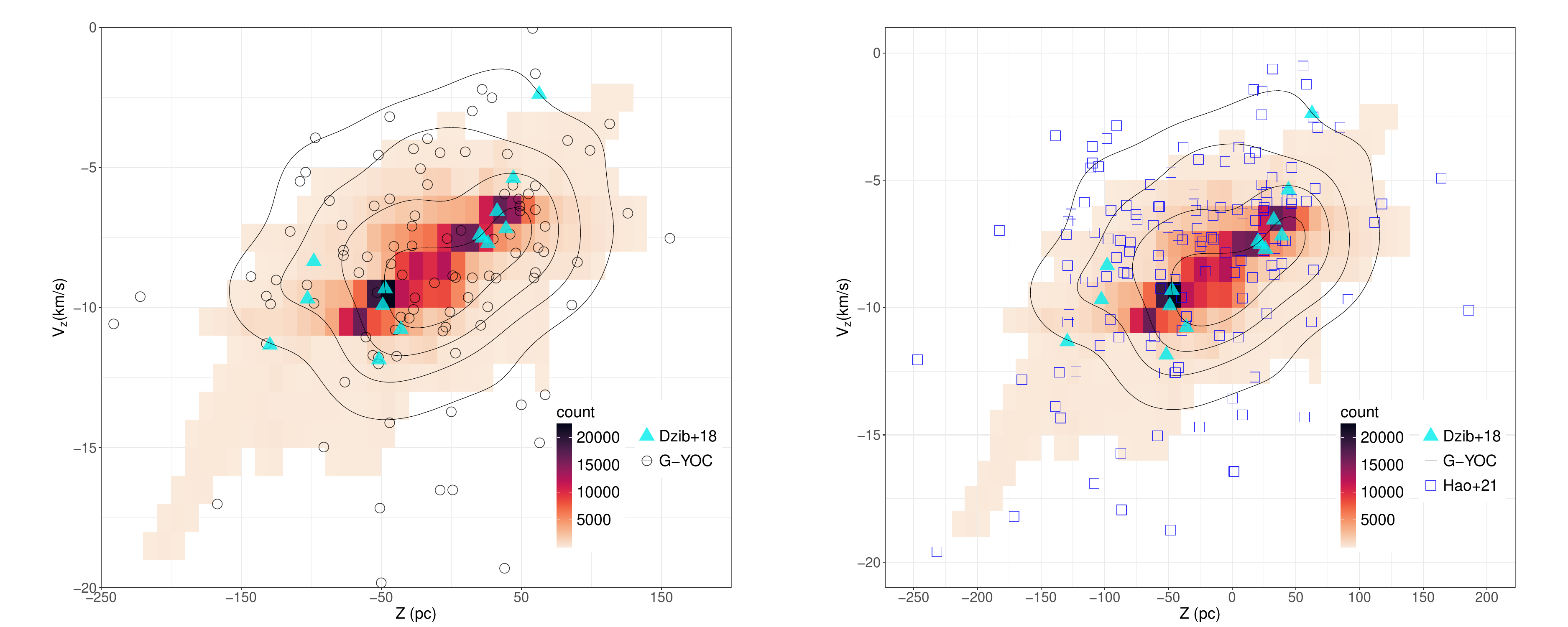}
    \caption{{ V$_Z$-Z diagrams obtained from the Kriging maps for the G-YOC sub-sample. Black circles and contour lines represent the distribution of G-YOC sample points, assessing the model’s fit quality. Star-forming regions (SFRs) close to the Sun from \citep{Dzib2018} are plotted as cyan-filled triangles. Additionally, a sub-sample of OCs from \citep{hao2021}, selected using the same criteria for age and distance as those applied to D-YOC and G-YOC, is represented by blue squares.}
    }
    \label{fig:Kriging2b}
    \end{subfigure}
    \caption{{Kriging maps for the OC sub-sample with $log(age) < 7.5$ \citep{Alfaro22}, with G-YOC shown in orange as heatmaps of 2D bins ($10$ pc by $1$ \kms in size) to approximate the surface density of points. Various OC samples from the literature, including the original one, are plotted on these maps to evaluate their consistency with the Kriging model’s data distribution.}}
    \label{fig:Kriging2}
\end{figure*}

\subsection{The observational vertical phase diagram}\label{subsec: method}

The young Galactic disc's spatial and kinematical vertical maps, $Z(X, Y)$ and V$_Z$(X, Y), are estimated by the Kriging method, an extensive family of spatial estimation methods for multi-dimensional stochastic processes. It optimizes existing knowledge by exploring how a target variable varies in space through a variogram model, considering spatial correlation, and providing the best linear unbiased estimator (BLUE) \citep{Oliver201456}. Although a more extended explanation of this mathematical tool can be found in the previous work \citep[][and references within]{Alfaro22}, let us briefly review this technique. 

For the open cluster (OC) samples we have $\left\{(\mathbf{r}_{i},Z(\mathbf{r}_{i}), V_Z(\mathbf{r}_{i}))\right\}_{i=1,\ldots,n}$  measurements of the target variables, i.e. vertical distance $Z(\mathbf{r}_{i})$ and vertical velocity $V_Z(\mathbf{r}_{i})$, at a certain number of points or locations $ \mathbf{r}_{i}(x_{i}, y_{i})$, for $i = 1,\ldots,n$, within a radius of $3.5$ Kpc approximately around the Sun. Note that we will only refer to $Z(\mathbf{r}_{i})$ in the following formulation for simplicity and clarity.

Suppose we want to obtain a spatial map of such magnitudes, estimated from a discrete set of measurements at different positions in the plane $\mathbf{r}_{i}$. In that case, we need to estimate the value of this variable $\hat{Z}(r)$ at any spatial coordinate $r$. This is based on the sample data and some assumptions about the formula trend of $Z(r)$, its variance, and some spatial correlation underlying the sample data. Regarding $Z(r)$ as a random process, it can be decomposed as 
\begin{equation}
\label{eq:krig_hyp}
Z(\mathbf{r}) = m + e(\mathbf{r}) 
\end{equation}
Where $ e(\mathbf{r}) $ is a zero mean random process with known covariance $ C(\mathbf{h}) = \text{E}[e(\mathbf{r})e(\mathbf{r} + \mathbf{h})] $,  $ \mathbf{h}(\mathbf{r}_{i}, \mathbf{r}_{j}) $ is the vector separation between pairs of sample points, and  $ m = \text{E}[Z(\mathbf{r})] $ is the mean of the process, which varies spatially, does not depend on $\mathbf{r}$ under the assumption of stationarity, and it can be modelled as linear function (so-called \textit{Universal Kriging}, UK) as 
\begin{equation}
\label{eq:krig_model}
m = \sum_{j=0}^p f_j(r)\beta_j = {F\beta} 
\end{equation}
where $f_j(r) = (f_j(r_1),\ldots,f_j(r_n))$ are $p$ known predictors or spatial regressors evaluated at each observation $r_i$, and ${F} = (f_1(r),\ldots,f_p(r))$.
$\mathbf{\beta} = (\beta_1,\ldots,\beta_p)^T$ are $p$ unknown regression coefficients. 
Given this model, the Kriging prediction of $Z(r)$ is the best linear unbiased predictor, given as
\begin{equation}
    \label{eq:krig_pred}
    \hat{Z}(r_0) =  f(r_0)\hat{\beta} \  + \  v^T V^{-1}(z(r)-F\hat{\beta})
\end{equation} 
The first term is the estimated mean value for location $r_0$, where $f(r_0) = (f_1(r_0),\ldots,f_p(r_0))$, and 
$\hat \beta = (F^T V^{-1} F)^{-1}F^T V^{-1}z(r)$ the generalised least squares estimate of the trend coefficients $\beta$.
The second term is a weighted mean of the residuals from the mean function, where $v^T V^{-1}$ are known as the Kriging weights, 
$v = Cov(Z(r_0), z(r))$, 
$V = Cov(e(r))$ is the \textit{known} covariance matrix of $z(r)$, 
and $z(r) = (Z(r_1),\ldots,Z(r_n))^T$. 
The prediction variance, or Kriging variance, is computed as 
\begin{eqnarray}
    \label{eq:sigma_krig}
    \sigma^2_{Z}(r_0) &= &Var(Z(r_0)-\hat{Z}(r_0)) = 
     \\
    && Var(Z(r_0)) \ \  - \ \ v^TV^{-1}v \ \ +  \nonumber \\
    && \scalebox{0.9}{$\left(f(r_0) - v^TV^{-1}F\right) (F^T V^{-1} F)^{-1} \left(f(r_0) - v^TV^{-1}F\right)^T $}\nonumber
\end{eqnarray}

Besides the stationarity assumption, that implies $ m = \text{E}[Z(\mathbf{r})] $ does not depend on $ \mathbf{r} $ and $ C(\mathbf{h})$ only depends on the distance (isotropy), it is also assumed that $ \text{E}[Z(\mathbf{r} + \mathbf{h}) - Z(\mathbf{r})] = 0 $. This is known as the intrinsic model, 
a weak hypothesis required in many practical situations for computing the variance of the process that yields 
\begin{equation}
    \label{eq:variance}
    \text{Var}[Z(\mathbf{r} + \mathbf{h}) - Z(\mathbf{r})] = 2\gamma(h) 
\end{equation}

Where $\gamma(h)$ is the variogram of the random process, and it is assumed to measure the spatial correlation in the actual realisation of $Z(\mathbf{r})$. Therefore, its computation is a crucial step for a correct or reliable estimation since it plays an essential tool providing the spatial dependence of spatially correlated variables, and it lets us estimate the value of the spatial variable at an un-sampled location,
\begin{equation}
\label{eq:variogram}
\hat \gamma(\mathbf{h}) = \frac{1}{2p(\mathbf{h})} \displaystyle \sum_{j = 1}^{p(\mathbf{h})} \Big\{ Z(\mathbf{r}_{j}) - Z(\mathbf{r}_{j} + \mathbf{h}) \Big\}^{2} \;  
\end{equation}
\noindent being $ p(\mathbf{h}) $ the number of point pairs separated by a distance $ \mathbf{h} \in \mathbb{R} $ (in the particular case under isotropy). 
The Appendix \ref{sec:The Variogram model} explains in more detail its estimation and model selection from the cross-validation between several theoretical variogram models. 
We mainly chose the spherical models for both variables, $Z$ and V$_Z$. Only in a few cases is the exponential model slightly better. 

Once the variogram estimator assessment $\hat \gamma(\mathbf{h})$ is done, providing the spatial dependency or auto-correlation in our data, we can finally perform the last stage of the Kriging approach to use the data to make {\it predictions} with equation  \eqref{eq:krig_pred}, letting us create a continuous surface of the phenomenon, as we can see at the left panels of  Fig. \ref{fig:Kriging1a}. 
{
These estimations were performed on a spatial grid encompassing the sampled area, specifically a region spanning 7 kpc around the Sun, divided into steps of 10 pc along each axis. The calculations were based on the fitted semivariogram model, \eqref{Eq:Sph_VM}-\eqref{Eq:Wave_VM}, and the spatial arrangement of the nearest input data. This approach results in a modelled Galactic plane surface grid comprising $\ 700\times700$  points, with the corresponding estimated values for $ Z$  and  $V_Z$ , respectively.}

Fig. \ref{fig:Kriging1a} presents the vertical phase diagrams for the four age groups of the D-YOC sample (in {blue-green} colour point density maps) and G-YOC data (in orange) previously computed in \citep{Alfaro22}. The age intervals have been chosen so that the central age value shows a low dispersion and can be assimilated into a coeval sample. The D-YOC sample was divided into four sub-samples by age to analyse the properties of the linear relationship between these two variables with age and, in particular, to estimate the sample's age, after which the linear pattern disappeared. We can see that the sample aged between 30 and 50 Ma shows no linear structure and seems to depict an incipient spiral pattern, as observed in older stars \citep{Antoja18,Antoja23}.

Table \ref{tab:vertical phase summary} summarises the sub-samples used in our study, characterised by their age. 
It shows the main statistics of the original sub-samples, the Kriged estimated $Z(X,Y)$ and $V_Z(X,Y)$ maps, and the linear fit of the $(V_Z,Z)$ distributions.
As we increase the central value of the age or the time interval of the sample, the dispersion of the ($V_Z$,Z) linear distribution increases. The linear relationship disappears for the oldest age group (log(t) between 7.5 and 7.7), and a quasi-elliptical, or spiral,  shape is drawn by the densest zones. The diagrams show the linear fit to the distribution as a solid line. This fit has been obtained with a weight proportional to the surface density in the vertical phase space.  

Comparing the vertical phase for age group $log(t) \leq 7.5$ of both samples (G-YOC and D-YOC), it is clear that the dispersion of the distribution is higher for the D-YOC sample. This result should not be surprising as there are zero-point and scale differences between the two samples in both age and distance variables \citep{Dias2021}. The most conspicuous age differences are found for the youngest clusters. However, the slope of the vertical phase space distribution is very similar for both diagrams. Two observational conclusions follow: the vertical phase space shows a linear shape for objects younger than 30 Ma, and the distribution slope decreases with age until $\sim$ 40 Ma, where the linear structure evolves into more complex shapes. Very young objects (age $\leq$ 30 Ma) show a linear vertical phase pattern, while the disc stars, with an average age older than 1 Ga, display a snail-shell structure \citep{Antoja23}.  

{Moreover, superimposed on the linear distribution observed in the $Z-Y$ and V$_Z$-Y maps, characterised by a higher point concentration (middle panels of Fig. \ref{fig:Kriging1a} and Fig. \ref{fig:Kriging2a}), there is a wave-like structure of greater amplitude with an apparent wavelength of approximately 1 kpc. Included in this wave-like structure are the Gould Belt young clusters within 500 pc of the Sun shown as cyan-filled triangles in Fig. \ref{fig:Kriging2a} \citep[from][]{Dzib2018}. This behaviour is unsurprising, as vertical corrugations along the  Y-axis are well-documented. These corrugations originate from the undulations observed in several spiral arms, whose predominant orientation runs parallel to the  Y-axis in this region of the Galactic plane. The existence of such corrugated arms has been recognised for over 50 years \citep[see][among others]
{Dixon1967, VQ1970, Quiroga1974, Lockman1977, SF1986, Alfaro1991, Alfaro1992} and has been interpreted through various theoretical frameworks \citep[][and references therein]{Nelson1976, Weinberg1991, Alfaro1996, Bland-Hawthorn21, Tepper2022}. Nevertheless, this study concentrates on the zero-order terms, representing the linear correlations between Z-Y and V$_Z$-Y, resulting in a linear distribution in the   V$_Z$-Z plane.}

Assessing how well the initial data aligns with the surfaces provided by Z(X,Y) and $V_Z(X,Y)$ Kriging estimation is crucial for validating the robustness of our methodology. We take the matrix data (X, Y, $\hat{Z}$, $\sigma_Z$, $\hat{V}_Z$, $\sigma_{V_Z}$) as the best representation of our disc.  
{Fig. \ref{fig:Kriging2}, similar to Fig.12 of \cite{Alfaro22}, shows the V$_Z$-Y, Z-Y, and V$_Z$-Z diagrams, where the colour maps represent the density of points on the diagram. We plotted the sample G-YOC (black points) with velocity and position data in this image. We have drawn the surface density contours of these objects (dark lines) superimposed in the last diagram to enhance the linearity given by the denser regions. In fact, despite the apparent outspread, the lighter colour corresponds to density values below the 10th percentile of the density points distribution.  Moreover, these diagram maps clearly show that the sample point density maxima corresponds to the estimated point density maxima. In other words, the V$_Z$-Z relation derived from the models not only represents the sample point distribution but also highlights the global behaviour of the Galactic disc. If we draw on this same plot the star-forming regions within a radius of 500 pc used by \cite{Dzib2018} to analyse the Gould Belt, we see that the points (green triangles) align perfectly with our map's regions of maximum density.}

{To provide a clearer understanding of the uncertainties associated with deriving the slopes $m({Vz}/{Z})$ from the Kriging maps, we include Fig. \ref{fig:Kriging2b}, which shows the $V_Z-Z$ relationship obtained for the G-YOC data. 
The left panel displays the G-YOC sample points (black circles) overlaid on the distribution derived from the Kriging maps. The pixel colour represents the surface density of points, with a bin size of $2.5$ pc in $Z$ and $0.25$ km s$^{-1}$ in V$_Z$.  
Despite the dispersion of the sample points, the areas of the highest cluster concentration coincide with the density maxima of the Kriging maps. Additionally, we include (green triangles) the star-forming regions associated with the Gould Belt \citep{Dzib2018}, which are constrained to within a 500 pc radius around the Sun. In the right panel, we incorporate a new sample of clusters selected from the \cite{hao2021} catalogue using the same criteria as for G-YOC. The distribution of these clusters follows a pattern consistent with the G-YOC sample, with the highest concentration located at the same position as the density maxima in the Kriging maps. These plots indicate that the $V_Z-Z$ distribution derived from the Kriging maps provides a reliable and robust representation of the Galactic plane as defined by the sample clusters.}

\begin{table*}
    \caption{Summary of the sample data for different age ranges, their estimated kriging maps Z(X,Y) and $V_Z(X,Y)$, and the vertical phase slope linear fits. }
    \scalebox{0.8}{
    \begin{tabular}{r r r|
    r@{$\ \pm \ $}l c 
    r@{$\ \pm \ $}l c 
    r@{$\ \pm \ $}l c
    r@{$\ \pm \ $}l 
    r}  
    \multicolumn{6}{c}{ Summary of  observations}  &
    \multicolumn{6}{c}{ Summary of  Kriging estimations}  &
    \multicolumn{3}{c}{Fitted Slope} \\
    \cmidrule(lr){1-6}\cmidrule(lr){7-12}\cmidrule(lr){13-15}
    \multicolumn{3}{c}{{Subset}} & 
    \multicolumn{3}{c}{Age (Ma)}  &
    \multicolumn{3}{c}{ Z (pc)}  &
    \multicolumn{3}{c}{V$_Z$ \ \  (km \,  s$^{-1})$}  &
    \multicolumn{2}{c}{m} & 
    $R^2$ \\
    \cmidrule(lr){1-3}\cmidrule(lr){4-6}\cmidrule(lr){7-9}\cmidrule(lr){10-12}
   $\#$ & Data & $log(t)$ & 
    $<t>$&$\sigma_t $ & $P_{10}-P_{90}$ &
    $<Z>$&$\sigma_{Z}$ & $IQR$ &
    $<V_Z>$&$\sigma_{V_Z}$ & $IQR$ &
    \multicolumn{2}{c}{$pc^{-1}km \,  s^{-1}$} 
    &  (\%) \\
    \hline
     1 & DYOC & $< 7$ & 
     $7.40$ & $ 1.52$ & $(5.48,9.62)$ & 
     $-25.25$ & $34.26$ & $(-43.80, -5.57)$ &
     $-8.11$ & $ 1.80$ &  $(-9.71,-6.62)$ & 
     $0.104$ & $ 2.04e-2 $ 
     &	$98.67$   \\[1ex]
     2 & DYOC & $7 - 7.5$ & $17.15$ & $ 6.01$ &  $(11.56 , 26.79)$ & 
     $-20.299$ & $40.51$ &  $(-49.22,7.27)$ & 
     $-7.28$ & $ 1.55$ &  $(-8.41,-6.16)$ & 
     $0.045 $ &  $1.03e-2$ & 	
     $78.26$   \\[1ex]
      3 &  DYOC & $< 7.5$ & $13.95$ & $ 8.28$ &  $6.338 - 24.95$ & 
      $-20.83$ & $39.70$ &  $(-46.99,5.49)$ & 
      $-7.78$ & $ 1.73$ &  $(-9.15,-6.38)$ & 
      $0.046 $ &$6.75e-3$&		
      $47.71$   \\[1ex]
     4 &  GYOC & $< 7.5$ & $17.66$ & $ 6.78$ &  $6.31 - 31.62$ &  
     $-17.54$ & $45.98$ &  $(-49.21,14.64)$ & 
     $-8.52$ & $ 1.62$ &  $(-9.73,-7.20)$ & 
     $0.035 $ &$1.16e-3$&%
     $99.3$   \\[1ex]
      5 &   DYOC & $7.5 - 7.7$ & $40.78$ & $  5.22$ &    $33.20-46.92$ & 
      $-3.96$ & $23.05$ &  $(-21.15,12.16)$ & 
      $-7.28$ & $ 1.56$ &  $(-8.41,-6.16)$ & 
      $-9.78e-4$ & $2.65e-3$ &	
      $0.05$  \\
      \hline
    \end{tabular}}
    \label{tab:vertical phase summary}
\end{table*}

\section{The vertical acceleration in a young, thin disc: comparison with observational vertical phase diagrams }\label{sec: Z_vs_Vz_theo}

The sample of clusters used in this work is aged less than $log(t) \leq 7.7$ and defines a young, thin disc with $|Z|$ values less than 200 pc. For this range of $Z$, and assuming a constant rotational velocity in the region and a steady and axisymmetric potential, it is satisfied that: a) the vertical acceleration  $K_Z$ depends only on $Z$ since the in-plane motions are decoupled from the vertical motion, and b) $K_Z \approx -4\, \pi\, G\,\rho_0\, Z$, which follows from the integration of Poisson's equation under the boundary conditions defined by our cluster sample. $\rho_0$ is the effective gravitational mass density representative of the analysed disc volume \citep[i.e.][]{Bahcall1984, Binney1987}.
This linear relationship between the vertical acceleration and the vertical distance to the Galactic plane is corroborated by previous dynamic studies of the thin Galactic disc for $|Z|\leq\,$ 200 pc \citep[i.e.][]{Siebert03, Salomon20}.

Under these assumptions, we can write 
\begin{equation}
    K_Z = -\Omega^2 \, Z
    \label{eq:Kz}
\end{equation}

\noindent  where $\Omega^2 = 4\pi G \rho_0$, with $G = 4.3009\times 10^{-3} $ pc M$_{\odot}^{-1}$ {km}$^2$  s$^{-2}$  and $\rho_0$ given in M$_{\odot}$ pc$^{-3}$. Solving \eqref{eq:Kz} we get the motion equations as: 
\begin{eqnarray}
    Z(t) & = & Z_0 \cos(\Omega t) + \frac{V_0}{\Omega} \sin (\Omega t)  = A \cos (\Omega t + \phi) 
    \label{eq:motZ} \\[1ex]
    V_Z(t) & = & V_0 \cos(\Omega t) - Z_0\Omega \sin (\Omega t)  = - A \Omega  \sin(\Omega t + \phi) 
    \label{eq:motVz}
\end{eqnarray}
where the time $t$ is measured in Ma, Z(t) in pc, and V$_Z(t)$ in km s$^{-1}$.  
Besides, at any time $t,$ it holds that
\begin{eqnarray}
    \label{eq:A2}
    Z^2(t) + \frac{V^2_Z(t)}{\Omega^2} 
     &=& A^2 \\ 
     \frac{V_Z(t)}{Z(t)} &=& - \Omega \tan(\Omega\, t + \phi)
\end{eqnarray}

The amplitude and phase (A,$\phi$) are related with the constants of the motion equations \eqref{eq:motZ}-\eqref{eq:motVz} as follows, 
\begin{eqnarray}
 Z_0^2 + \frac{V_0^2}{\Omega^2} = A^2  
 \quad , \quad
\frac{V_0}{Z_0\Omega} = - \tan(\phi) 
\end{eqnarray}

One way to look at equations (\ref{eq:motZ}) and (\ref{eq:motVz}) is to consider a distribution of coeval clusters synchronized in the Y direction, i.e., which is mostly along the local spiral arm, that drift vertically from the same initial position, $Z(0)$, starting at the same time, but with slightly different velocities, V$_Z$(0). Over time, these clusters will spread out in Z, with the initially faster ones moving to greater Z. For example, if they all start in the mid-plane with negative V$_Z$, then $\phi$=$\pi/$2. The initial dispersion in vertical velocity is represented by a dispersion in $A$, $\Delta V_Z=\Delta A\Omega$, and $\Delta Z=0$ initially. After a time $t$ and they spread out, the tilt of their distribution on a $(V_Z,Z)$ plane is 
\begin{equation}
    {\frac{\Delta V_Z}{\Delta Z}} = 
-\Omega\cot(\Omega t).
\end{equation}
For $\Omega t<<1$ this ratio is $-1/t$, and
the extent of their height is $\Delta Z(t)=\Delta A\Omega t$.

The slopes of the cluster complexes on the $(V_Z,Z)$ plane, m, are given in Table 1. In this example, the product of the average cluster age and the slope should equal $\Omega$ t $\cot$($\Omega$ t) with Z$_0$=0. For representative m=0.03 pc$^{-1}$ \kms and $<t>\sim10$ Ma,  m<t> $\sim$ 0.3, and then $\Omega$ t is solved to be $\sim$ 1.3. Dividing this by the age gives $\Omega\sim$0.13 Ma$^{-1}$, and a corresponding average mid-plane density $\rho_0\sim 0.3$ M$_\odot$ pc$^{-3}$.
These volumetric density values, computed as above, are higher than most of the calculated by other authors \citep[e.g.][]{Binney1987, 1974Natur.252..272H, 1974ApJ...191L..57V}. 

\begin{figure}
\includegraphics[width=0.45\textwidth]{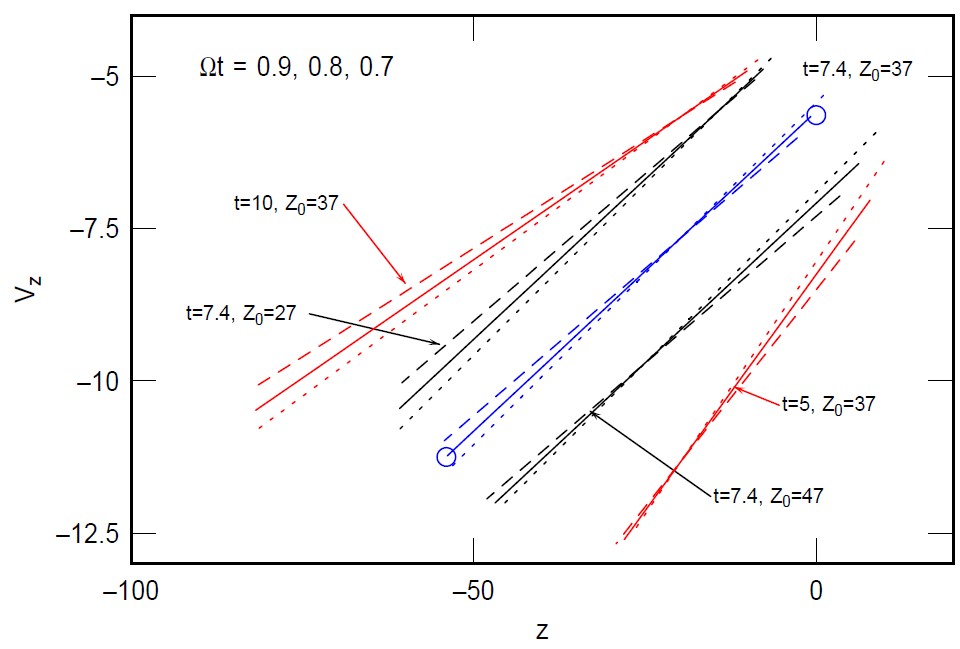}
\caption{Several solutions of equations (\ref{eq:motZ}) and (\ref{eq:motVz}) with slightly varying parameters, $\Omega t$, $t$ in Ma, $Z_0$ in pc, and $V_0$ varying along the curves in the range from $-4$ \kms  to $-12$ \kms .}
\label{fig:orbits}
\end{figure}

A detailed fit to the distribution of cluster velocities and positions requires both initial positions and velocities. Fig. \ref{fig:orbits} shows several solutions of equations (\ref{eq:motZ}) and (\ref{eq:motVz}) with slightly varying parameters, including a reasonable fit with $\Omega t=0.8$, $t=7.4$ Ma, $Z_0=37$ pc, and $V_0$ in the range from $-4$ \kms to $-12$ \kms, which goes through the circles representing the main peaks in  Fig. \ref{fig:Kriging1a}. The different clusters of curves are for different ages (lower ages to the lower right, higher ages to the upper left), and different starting positions (higher positive $Z_0$ to the lower left and lower positive $Z_0$ to the upper right). For each cluster of curves, the dashed, solid and dotted lines are for $\Omega t=1,$ 0.9, and 0.8. The best fit with $\Omega t=0.9$ and $t=7.4$ Ma implies a mid-plane density of $0.22$ M$_\odot$ pc$^{-3}$. 

From the equation \eqref{eq:A2}, V$_Z$ can be rewritten as a function of Z(t), and we can derive the function slope at any time t as,
\begin{eqnarray}
    m(t) = \frac{\partial\,V_Z(t)}{\partial\,Z(t)} 
    &=& 
    \frac{\Omega}{\tan(\Omega\, t + \phi)}   
    \ \quad \text{(pc$^{-1}$ \kms )}  
    \label{eq:slope}
    \\[1ex]
    &=& \Omega \frac{1-\tan(\Omega\, t )\tan(\phi)}{\tan(\Omega\, t ) + \tan(\phi) }
    \label{eq:slope2}
    \\[1ex]
    &=&  \frac{ \Omega-\tan(\Omega\, t )\frac{V_0}{Z_0}}{\tan(\Omega\, t ) + \frac{V_0}{Z_0\Omega}}
    \label{eq:slope3}
\end{eqnarray}
We have that the slope of the vertical phase diagram of a set of coeval objects born near the Galactic plane satisfies the conditions for their vertical motion to obey the equation \eqref{eq:Kz}. In other words, stars or clusters of age t, born at a similar oscillatory phase ($\phi$),  with $|$Z$_0|<<$ 200 pc, will form a straight line in the (V$_Z$, ) diagram with a slope given by equation \eqref{eq:slope}. 

\subsection{Model-Observations Comparison}

\begin{figure*}
\includegraphics[width=\textwidth]{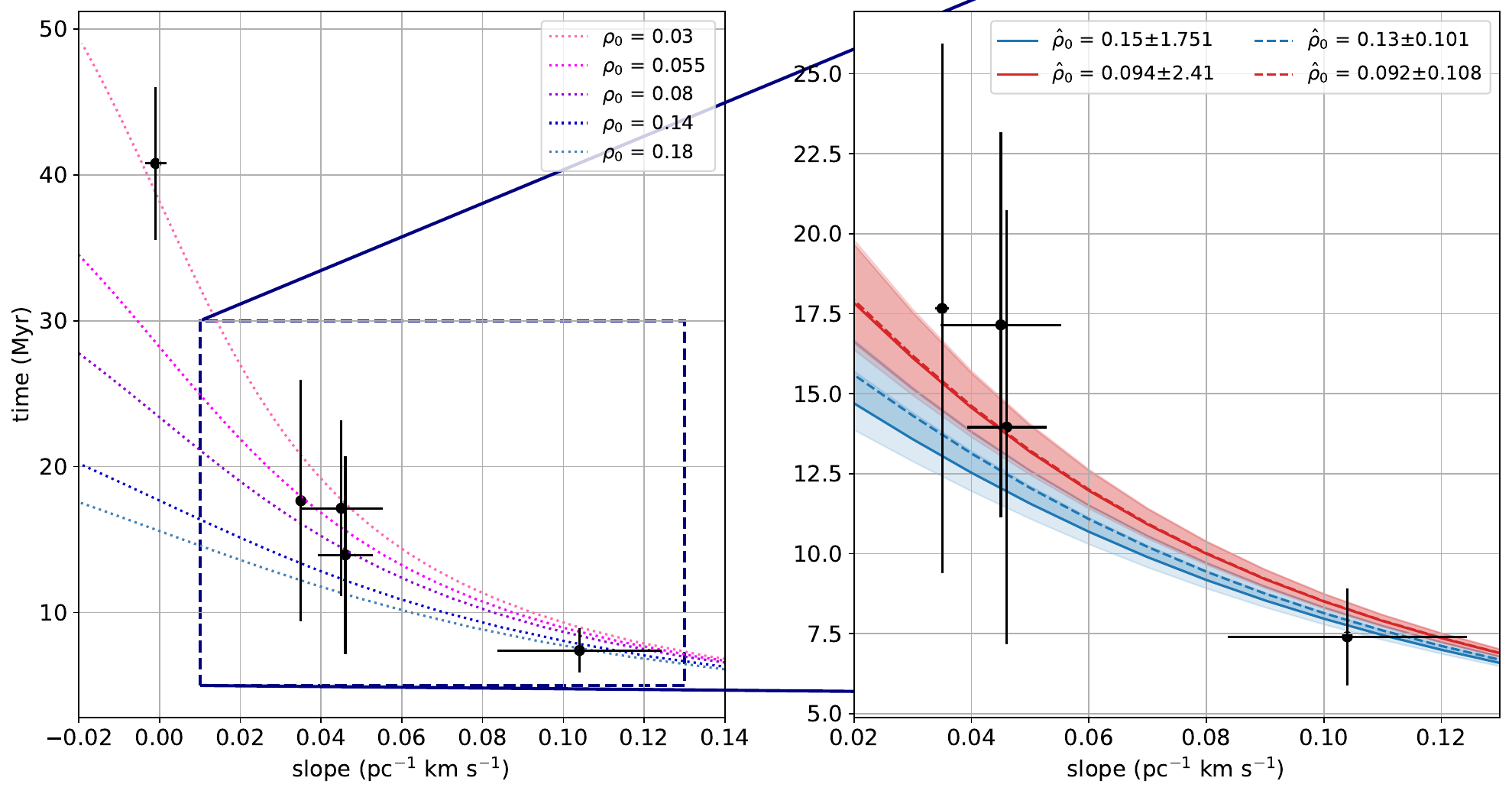}
\caption{\textit{At the left} Models versus data comparison of the slope between $V_Z$ and $Z$, $m(t)$ \eqref{eq:slope}. The different averaged stellar density $\rho_0$ values for the mid-plane are examined from the literature. The black points represent the estimated $\hat m (t)$ for the YOC samples, together with their mean ages (see Table \ref{tab:vertical phase summary});
\textit{At the right} The model estimation or fit derived from our data points, removing the older stellar population, i.e., the $7.7 \leq log(age)\leq 7.7$ sub-sample. In blue, the corresponding fits to the four points: non-weighted (continuous line) and weighted (dashed line). Similarly, in red for the 3 points fit. The shaded colours represent the RMSE from the different models. }
\label{fig:Z_VS_Vz_models}
\end{figure*}

The slope of the (V$_Z$, Z) diagram is a function of time $t$ and phase $\phi$, which implicitly depends on the initial conditions and the effective gravitational mass density in the region. Only for a set of coeval objects at the same oscillatory phase $\phi$ can a linear distribution be observed in the vertical phase diagram, and yet this is what we see in the (V$_Z$, Z) planes defined by the younger clusters. At the left panel of  Fig. \ref{fig:Z_VS_Vz_models}, we show m(t)  for five different values of $\rho_0$ with $\phi=$0. The estimated values of m(t) for the different age groups are superimposed in the plot. 

The D-YOC $V_Z - Z$ diagram for $log(t)$ between 7.5 and 7.7 deviates from linearity, and the correlation coefficient is almost zero. The central value of the age is 40 Ma, which seems to define the time limit for which the linear distribution degrades and a more structured pattern appears. Thus, given the observed distribution, the slope obtained does not represent a horizontal line but a low correlation coefficient.

At the left panel of  Fig. \ref{fig:Z_VS_Vz_models} we represent the model slopes \eqref{eq:slope} for three different theoretical stellar densities $\rho_0$, based on the dynamical mass density in the Solar neighbourhood estimated in the range 0.09 to 0.12 $\text{M}_{\odot}\, \text{pc}^{-3}$ by \citet{Binney1987},  $0.14 \ \text{M}_{\odot}\, \text{pc}^{-3}$ by \cite{1974Natur.252..272H}, or up to $0.18$ by \cite{1974ApJ...191L..57V}. We added the smallest values, $\rho_0=0.03$ and $\rho_0=0.055$, for illustrative purposes to check the required value to achieve the oldest sub-sample of OCs. The latter is indeed excluded from fitting this parameter, as shown in the right panel of Fig. \ref{fig:Z_VS_Vz_models} since their $V_Z-Z$ relationships departs from linearity as already discussed above. 
The black points represent the different slopes measured for the four sub-samples from D-YOC and G-YOC samples, and the error bars consider both the Kriging and model fitting uncertainties. Table \ref{tab:vertical phase summary} shows these fits. 

The $\rho_0$ estimates, although with a large uncertainty given the limited number of points for a non-linear fit, reproduce the values published by other authors. This agreement seems to support the internal consistency of our dynamic model. %

\begin{table}
    \centering
    \caption{Summary of the parameters estimation from model \eqref{eq:slope}.}
    \begin{tabular}{c|c r@{$\ \pm \ $}l c }
    No. of samples & weights & \multicolumn{2}{c}{$\hat{\rho}_0$} & RMSE \\
    \hline
         \multirow{2}{*}{4}& $none$ &$0.15$ &$1.75$ & $0.0377$\\[1ex]
          &$\sigma_{t}^{-1}$ & $0.13$ & $0.101$& $0.0377$ \\[1ex]
          \hline
         \multirow{2}{*}{3}& $none$ &$0.094$ &$2.4$ & $0.0313$\\[1ex]
         &$\sigma_{t}^{-1}$ & $0.092$ & $0.108$& $0.0313$ \\
    \hline
    \end{tabular}
    \label{tab:fitted_rho}
\end{table}

\section{Discussion}

Our study begins with a sample of young clusters, log(age) $< 7.7$, within a square $(X,Y)$ of 7 kpc on each side from a catalogue with well-determined distances. This careful selection ensures the reliability of our data from the outset. We consider these objects to be samples of the 3D structure of the young disc, which is a discrete collection of 3D objects.

Applying the Kriging methodology, we obtain the BLUE (Best Linear Unbiased Estimator) surfaces ${Z}(X, Y)$ and ${V}_Z(X, Y)$ which represent our Galactic disc model. The fact that the distribution of sample points is not homogeneous only means that the estimate's uncertainty will be greater in the areas of lower sample density. 

This methodology also estimates the uncertainties $\sigma_Z$ and $\sigma_{V_Z}$ at each location $(X,Y)$ of the grid matrix, where $Z(X,Y)$ and $V_Z(X,Y)$ values have been computed. This uncertainty encompasses different aspects, such as the $Z$ height scale of the clusters' distribution, the spatial distribution of the sample, and the sampling uncertainty of the variables $Z$ and $V_Z$ \citep[see][for more details]{Alfaro1991}. 

We recall this modelling is our best representation of the Galactic disc, and the matrices ${Z}(X,Y)$ and ${V}_Z(X,Y)$ set up the database on which we analyse the $V_Z-Z$ relationship.  

The spatial distribution of the initial data can be interpreted in different ways, either as three continuous spiral arms or as a discrete collection of star-forming regions, where each grouping contains other smaller ones in a cascade representative of hierarchical star formation, as we already discussed in the previous work \citet{Alfaro22} where we analysed the vertical structure along different lines on the plane and its comparison with observational data from other authors. We now consider a surface, not a discrete collection of lines or regions on the plane. 

For instance, a linear relationship also derives from $Z-Y$ and $V_Z-Y$, where $Z(Y)$ and $V_Z(Y)$ are obtained projecting $Z(X,Y)$ and $V_Z(X,Y)$ on the $Y$ axis collapsing the $X$ axis (see  Figs. \ref{fig:Kriging1a} and \ref{fig:Kriging2}). The Galactic quadrants I and II show more positive values than quadrants III and IV for both variables. This observational fact derived here for the young plane (age < 30 Ma) has been previously detected in other studies \citep[e.g.][]{Romero2019}. It appears to be a global phenomenon of the disc, on which small local variations are observed, which could represent bubbles, corrugations or other phenomenologies associated with active star-forming regions. {It is worth highlighting that the star-forming regions associated with the Gould Belt \citep{Dzib2018} align closely with the denser regions of the  $V_Z-Z$ diagram (cyan triangles in Fig. \ref{fig:Kriging2a}). However, this alignment does not extend to the larger-scale $V_Z-Y$ and  Z-Y  diagrams, where the Gould Belt data exhibit a significantly steeper distribution than the Kriging maps' main dataset.} 

\citet{2023ApJ...943...88L} have recently analysed the $V_Z-Z $ diagrams for three different stellar populations, finding roughly linear relationships between the two variables. These authors interpret these diagrams in terms of Galactic warp. Interestingly, the distribution of OB stars, with an average age similar to the D-YOC sample, presents a $V_Z-Z$ gradient similar to the one found in this work for $|Z|$ values less than 200 pc.  

We propose another interpretation for our young OCs data based on the simplest dynamical model, the harmonic oscillation of each object. For a group of coeval stellar objects subjected to this potential and born in phase, we expect a linear relationship between $V_Z-Z$, the slope of which depends on the age of the objects, the phase, and the effective mass density in that volume. As time progresses and the sample's age range increases, the slope of the linear relationship changes and becomes noisier until it reaches a certain age ($log(t) >7.5$), at which point the linear structure breaks down, giving rise to a quasi-elliptical distribution. We can consider that an indirect proof that this simple model explains the observations is that the volumetric mass densities obtained from this model are compatible with those obtained by other authors. However, we determine them with great uncertainty.

Another fact that constrains the problem is the Galactic potential, which best describes the physics of the system. Many previous studies consider a decoupling between in-plane motions and vertical motions under certain conditions. For $|Z| < 200$ pc, the vertical acceleration $K_Z$ is proportional to $Z$ \citep{Salomon20}, as long as we are in a region of the Galaxy where the rotational velocity can be considered constant. In other words, we can add whatever we want to the vertical Galactic potential, such as spiral arms, transient injections of momentum and energy, etc., but it is evident that the vertical harmonic potential is the basic ingredient.

Under this simple dynamical model, we can explain the linear correlation between $V_Z$ and $Z$ for coeval clusters and the slope variation with age. The values of $\rho_0$ driving the oscillation frequency are compatible with previous estimates obtained with different methods. 
We also find that $V_Z-Z$, for clusters older than 40 Ma, deviates from linearity, showing a density distribution that more closely resembles an emerging spiral or ellipse on the $V_Z-Z$ plane. For ages older than 40 Ma, the dynamic problem needs ingredients other than the simple harmonic oscillator. In particular, the linear distributions between $V_Z-Y$ and $Z-Y$ (Figure \ref{fig:Kriging1a}, middle vertical panel, and Figure \ref{fig:Kriging2a}, top right and central panel) cannot be simply explained by a  Galactic potential as $K_Z \approx -\Omega^2\ Z$ for clusters older than 40 Ma.
{Finally, we conclude that for ages less than 40 Ma, the most significant vertical driving force derives solely from the harmonic oscillator potential, generated by the in-plane matter distribution and characterised by an effective mass density $\rho_0$. The observed linear behaviour supports the hypothesis that clusters of similar age are formed with a comparable phase in this oscillation. The deviation from linearity observed in older clusters may be attributed to greater phase dispersion over time in these objects caused by transient phenomena. Such phenomena would introduce higher-order and non-stationary terms into the Galactic potential, disrupting the decoupling between the vertical phase and in-plane motion.
}

\section*{Acknowledgements}

{Helpful comments from the referee are gratefully acknowledged.}
This work has been partially funded by the Spanish MCIN/AEI grant PID2022-136640NB-C21.
EJA acknowledges financial support from the Severo Ochoa grant CEX2021-001131-S funded by MCIN/AEI/ 10.13039/501100011033. EJA  has made continuous use of TopCat \citep{TopCat2005}. This work presents results from the European Space Agency (ESA) space mission's Gaia data. Gaia data are being processed by the Gaia Data Processing and Analysis Consortium (DPAC). Funding for the DPAC is provided by national institutions, in particular, the institutions participating in the Gaia MultiLateral Agreement (MLA). The Gaia mission website is \url{https://www.cosmos.esa.int/gaia}. The Gaia archive website is \url{https://archives.esac.esa.int/gaia}.







\appendix 

\renewcommand\thefigure{\thesection.\arabic{figure}} 
\setcounter{figure}{0}

\renewcommand\thetable{\thesection.\arabic{table}} 
\setcounter{table}{0}

\section{Variogram fit and Model Selection}\label{App:VMS}

\subsection{The Variogram model}\label{sec:The Variogram model}

\begin{figure}
        \includegraphics[width=0.49\textwidth]{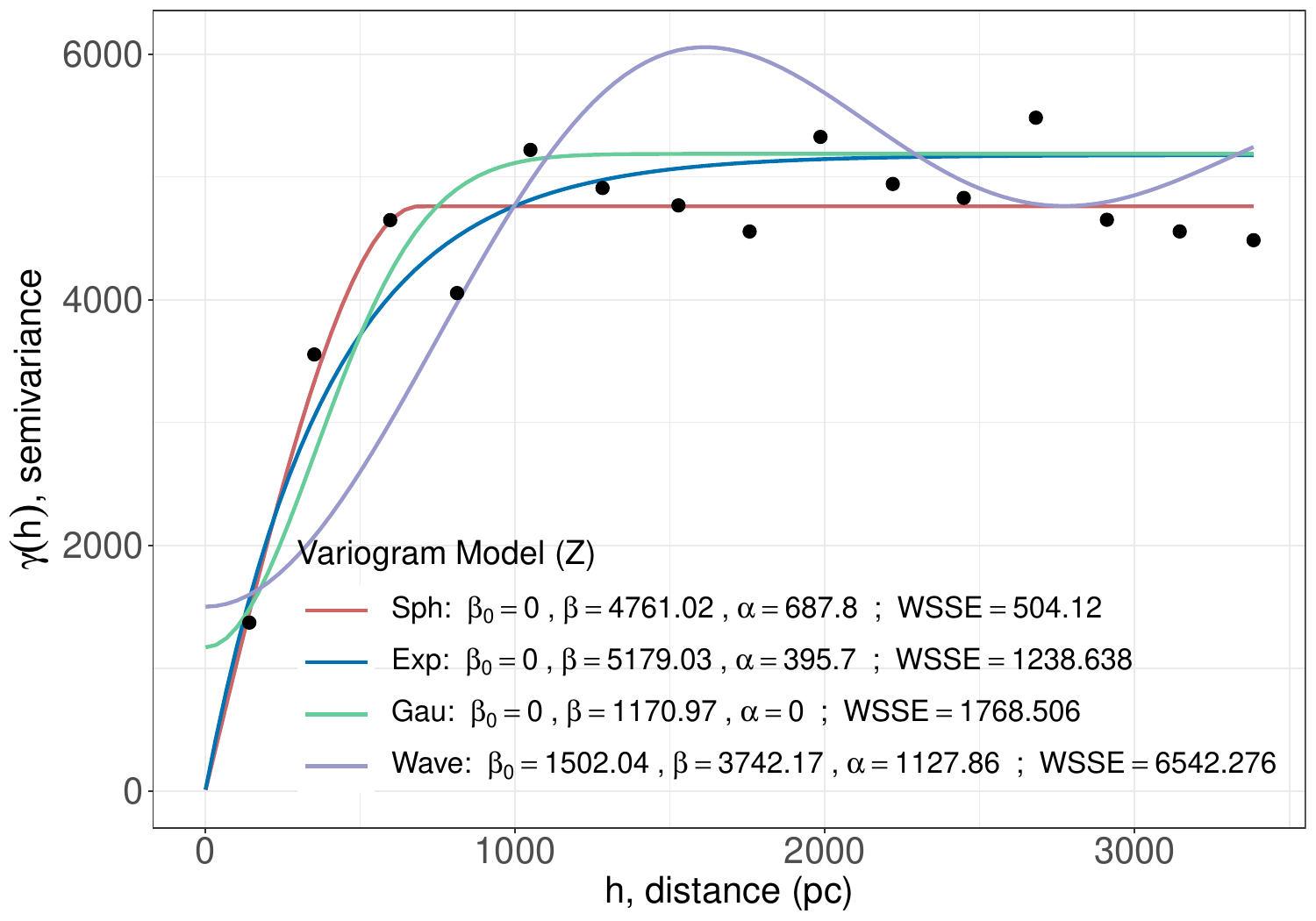} 
        \caption{$Z$ Variogram model fit for $log(age) < 7 $. }
        \label{fig:Z_7}
\end{figure}


Variogram estimation plays a key role in the Kriging method since it provides a useful tool for measuring the spatial dependence of spatially correlated variables and lets us estimate the value of the spatial variable at an un-sampled location. The Kriging estimation considers a weighted mean \eqref{eq:krig_pred}, such that the closer measurements of the un-sampled location are more closely related to the actual unknown value than further ones. The variogram estimation is commonly used to find the optimal values of these weights, and it requires the computation of the experimental variogram, or {\it empirical variogram estimation} $\hat \gamma(\mathbf{h})$, from the sample set $S$, computed as \eqref{eq:variogram}

The sample variograms and co-variograms are commonly calculated from predicted residuals 
$\hat{e}(r_i) = z(r_i) - \hat{m}(r_i)$, 
where $\hat{m} (r_i) = F\hat{\beta}$ is the ordinary least square estimates of the mean. So an equivalent to Equation \eqref{eq:variogram} is, 
\begin{equation}
\label{eq:variogram_bis}
\hat \gamma(\bar{h}_j) = \frac{1}{2N_j} \displaystyle \sum_{i = 1}^{N_j} \left( \hat{e}({r}_{i}) - \hat{e}({r}_{i} + {h}) \right)^{2}
\end{equation}
for all $(r_i,r_i+h)$ such that $h \in [h_j,h_j+\delta]$. This is for a number $N_j$ of pairs within a regular distance interval $[h_j,h_j+\delta]$, with $\bar{h}_j$ the average of such interval. %
We then get a variogram estimator by fitting the empirical variogram with some theoretical model, among some family of valid variograms that best fits the experimental variogram and captures the underlying spatial dependence of the data. 

We tested with the more extended semivariogram models: exponential, spherical, and Gaussian. We also tried using the wave model due to its atypical irregular behaviour, taking into account the expected corrugated nature of the $Z$ variable.
\begin{align}
        \gamma_{sph}(h) &=  
    \beta_0 + \beta \left( {3h}/{2\alpha}-\left({h}/{\alpha} \right)^3/2 \right),
        & 0 < h \leq \alpha
        \label{Eq:Sph_VM}\\
       & \quad \beta_0 + \beta, & h > \alpha 
    \nonumber\\[1.5ex]
    \gamma_{exp}(h)   &=  
    \beta_0 + \beta \left( 1 - exp\left({-{h}/{\alpha}}\right)\right), 
     & h > 0 
    \label{Eq:Exp_VM}\\[1.5ex]
    \gamma_{gau}(h)   &=   
    \beta_0 + \beta ( 1 - exp({-{h^2}/{\alpha^2}})), 
     & h > 0 
    \label{Eq:Gau_VM}\\[1.5ex]
    \gamma_{wave}(h)   &=  
    \beta_0 + \beta \left( 1 - {\alpha}/{h} \, sin\left(-{h}/{\alpha}\right) \right), 
     & h > 0
    \label{Eq:Wave_VM}
\end{align}
where $\beta_0$ is the {\it nugget}, which reflects local effects or sampling error; An important characteristic is that the semi-variance function $\gamma(h)$ can approach, or asymptotically converge to, a constant value known as the {\it sill}, $\beta_0 + \beta$; $ \beta$ is the {\it partial sill}. 
And $\alpha$ is the {\it range}, the distance the semivariogram reaches the sill. Locations separated by distances shorter than the range are spatially auto-correlated, while the correlation between locations beyond this distance is zero.


 Fig. \ref{fig:Z_7} shows the experimental variogram, black dots, for the spatial $Z$ component, fitted with the four theoretical models described above. As analysed below, we chose the exponential model as the best-fit model for the $Z$ semivariogram and the spherical one for the velocity component $V_Z$.

\subsection{Cross-Validation and Model Selection}

\begin{table*}
\caption{Variogram fits and cross-validation results.}
    \centering
    \scalebox{0.8}{
    \begin{tabular}{r rrrr@{$\ \pm \ $}lrr rrrr@{$\ \pm \ $}lrr}
       \toprule
      \multirow{2}{*}{\small $log(age) \leq 7$}  &
      \multicolumn{7}{c}{$Z$} & \multicolumn{7}{c}{$V_Z$} \\
       \cmidrule(lr){2-8}\cmidrule(lr){9-15} 
         & \multicolumn{1}{c}{WSSE} & 
         \multicolumn{1}{c}{AIC} & 
         \multicolumn{1}{c}{$R^2$} & 
         \multicolumn{2}{c}{$Z_{score}$} & \multicolumn{1}{c}{$\hat{Z} \sim Z_{obs}$} & 
         \multicolumn{1}{c}{$Z_{res} \sim \hat{Z}$ } &
         \multicolumn{1}{c}{WSSE} & 
         \multicolumn{1}{c}{AIC} & 
         \multicolumn{1}{c}{$R^2$} & 
         \multicolumn{2}{c}{$V_{Z_{score}}$}&
         \multicolumn{1}{c}{$\hat{V}_Z \sim V_{Z_{obs}}$} & 
         \multicolumn{1}{c}{$V_{Z_{res}} \sim \hat{V}_Z$}
         \\
         \hline
        Exponential  & $1238.64 $ & $-243.84$ & $99.960$ & $-0.086$& $1.048$ & $0.626$ & $-0.064$ & $0.041$ & $-121.40$ & $96.207$ & $-0.021$& $0.868$ & $0.259$ & $-0.090$ \\
        Gaussian  & $1768.51 $ & $-206.97$ & $99.958$ & $-0.020$& $1.253$ & $0.477$ & $-0.269	$ & $0.044$ & $-117.22$ & $96.210$ & $-0.029$& $0.853$ & $0.345$ & $-0.029$ \\
         Spherical  & $504.12 $ & $-242.73$ & $99.956$ & $-0.061$ & $1.203$ & $0.601$ & $-0.107$ & $0.040$ & $-122.32$ & $96.164$ & $0.010$ & $0.838$ & $0.349$ & $0.008$ \\
         Wave  & $6542.28 $ & $-243.08$ & $99.960$ & $-0.048$& $1.336$ & $0.448$ & $-0.210$ & $0.041$ & $-122.69$ & $96.290$ & $-0.004$& $0.808$ & $0.441$ & $0.078$ \\[0.5ex]
        \hline
      \multirow{2}{*}{\small $7 < log(age) \leq 7.5$}  &
      \multicolumn{7}{c}{$Z$} & \multicolumn{7}{c}{$V_Z$} \\
       \cmidrule(lr){2-8}\cmidrule(lr){9-15} 
         & \multicolumn{1}{c}{WSSE} & 
         \multicolumn{1}{c}{AIC} & 
         \multicolumn{1}{c}{$R^2$} & 
         \multicolumn{2}{c}{$Z_{score}$} & \multicolumn{1}{c}{$\hat{Z} \sim Z_{obs}$} & 
         \multicolumn{1}{c}{$Z_{res} \sim \hat{Z}$ } &
         \multicolumn{1}{c}{WSSE} & 
         \multicolumn{1}{c}{AIC} & 
         \multicolumn{1}{c}{$R^2$} & 
         \multicolumn{2}{c}{$V_{Z_{score}}$}&
         \multicolumn{1}{c}{$\hat{V}_Z \sim V_{Z_{obs}}$} & 
         \multicolumn{1}{c}{$V_{Z_{res}} \sim \hat{V}_Z$}
         \\
         \hline
        Exponential & $1339.22$ & $-246.64$ & $99.974$ & $-0.002$ & $1.038$ & $0.496$ & $-0.049$ & $0.002$ & $-53.59$ & $89.536$ & $0.019$ & $0.963$ & $0.344$ & $-0.097$ \\
        Gaussian & $1786.12$ & $-234.19$ & $99.975$ & $-0.010$ & $1.032$ & $0.510$ & $-0.011$ & $0.002$ & $-56.01$ & $89.465$ & $0.003$& $0.961$ & $0.353$ & $-0.062$\\
        Spherical & $1265.83$ & $-246.68$ & $99.973$ & $0.005$& $1.039$ & $0.512$ & $-0.037$ & $0.003$ & $-52.93$ & $89.229$ & $0.001$& $0.948$ & $0.405$ & $-0.004$\\
        Wave & $2137.54$ & $-246.30$ & $99.973$ & $0.001$& $1.111$ & $0.461$ & $-0.083$ & $0.002$ & $-52.90$ & $89.294$ & $0.007$ & $0.973$ & $0.351$ & $-0.093$\\[0.5ex]        
        \hline
      \multirow{2}{*}{\small $log(age) \leq 7.5$}  &
      \multicolumn{7}{c}{$Z$} & \multicolumn{7}{c}{$V_Z$} \\
       \cmidrule(lr){2-8}\cmidrule(lr){9-15} 
         & \multicolumn{1}{c}{WSSE} & 
         \multicolumn{1}{c}{AIC} & 
         \multicolumn{1}{c}{$R^2$} & 
         \multicolumn{2}{c}{$Z_{score}$} & \multicolumn{1}{c}{$\hat{Z} \sim Z_{obs}$} & 
         \multicolumn{1}{c}{$Z_{res} \sim \hat{Z}$ } &
         \multicolumn{1}{c}{WSSE} & 
         \multicolumn{1}{c}{AIC} & 
         \multicolumn{1}{c}{$R^2$} & 
         \multicolumn{2}{c}{$V_{Z_{score}}$}&
         \multicolumn{1}{c}{$\hat{V}_Z \sim V_{Z_{obs}}$} & 
         \multicolumn{1}{c}{$V_{Z_{res}} \sim \hat{V}_Z$}
         \\
         \hline
        Exponential & $1149.19$ & $-245.04$ & $99.967$ & $ 0.003$ & $ 1.109$ & $0.540$ & $ -0.086$ & $ 0.012$ & $ -69.97$ & $89.051$ & $-0.016$ & $ 1.041$ & $0.475$ & $ -0.050$ \\
        Gaussian & $2845.50$ & $-224.38$ & $99.967$ & $-0.001$ & $ 1.097$ & $0.539$ & $ -0.060$ & $ 0.023$ & $ -47.62$ & $91.660$ & $ 0.000$ & $ 0.943$ & $0.449$ & $ -0.005$ \\
        Spherical & $2549.44$ & $-244.75$ & $99.969$ & $-0.017$ & $ 1.090$ & $0.553$ & $ -0.056$ & $ 0.012$ & $ -69.74$ & $91.578$ & $ 0.001$ & $ 0.941$ & $0.489$ & $0.016$ \\
        Wave & $12278.13$ & $-244.81$ & $99.965$ & $-0.006$ & $ 1.204$ & $0.472$ & $-0.098$ & $ 0.061$ & $ -70.60$ & $90.749$ & $ 0.021$ & $ 1.016$ & $0.467$ & $ -0.079$ \\[0.5ex]
        \hline
      \multirow{2}{*}{\small $7.5 < log(age) \leq 7.7$}  &
      \multicolumn{7}{c}{$Z$} & \multicolumn{7}{c}{$V_Z$} \\
       \cmidrule(lr){2-8}\cmidrule(lr){9-15} 
         & \multicolumn{1}{c}{WSSE} & 
         \multicolumn{1}{c}{AIC} & 
         \multicolumn{1}{c}{$R^2$} & 
         \multicolumn{2}{c}{$Z_{score}$} & \multicolumn{1}{c}{$\hat{Z} \sim Z_{obs}$} & 
         \multicolumn{1}{c}{$Z_{res} \sim \hat{Z}$ } &
         \multicolumn{1}{c}{WSSE} & 
         \multicolumn{1}{c}{AIC} & 
         \multicolumn{1}{c}{$R^2$} & 
         \multicolumn{2}{c}{$V_{Z_{score}}$}&
         \multicolumn{1}{c}{$\hat{V}_Z \sim V_{Z_{obs}}$} & 
         \multicolumn{1}{c}{$V_{Z_{res}} \sim \hat{V}_Z$}
         \\
         \hline
        Exponential & $605.61$ & $-238.33$ & $ 99.972$ & $0.002 $ & $ 1.025$ & $0.185$ & $-0.185$ & $0.029$ & $-87.08$ & $ 95.833$ & $0.022$ & $ 0.812$ & $0.452$ & $0.029$\\
        Gaussian  & $622.39$ & $-227.08$ & $ 99.970$ & $0.006$ & $ 1.015$ & $0.254$ & $-0.129$ & $0.027$ & $-75.73$ & $ 95.579$ & $0.022$ & $ 0.859$ & $0.380$ & $-0.066$\\
        Spherical  & $257.23$ & $-237.54$ & $ 99.972$ & $-0.009$ & $ 0.950$ & $0.348$ & $-0.017$ & $0.016$ & $-84.88$ & $ 95.305$ & $0.022$ & $ 0.845$ & $0.439$ & $0.036$\\
        Wave  & $829.00$ & $-237.41$ & $ 99.969$ & $0.007$ & $ 1.093$ & $0.166$ & $-0.248$ & $0.045$ & $-79.88$ & $ 95.052$ & $0.039$ & $ 0.958$ & $0.299$ & $-0.031$\\[0.5ex] 
        \bottomrule 
    \end{tabular}}
    \label{tab:Table_AIC_WSSE}
\end{table*}

The model selection assessment is based on several statistical evaluations, summarised in Table \ref{tab:Table_AIC_WSSE}. One of them is the weighted sum of square errors from the model variogram fitting stage, 
\begin{equation}
    \label{eq:WSSE}
    WSSE = \sum_{j=1}^{n} \frac{N_j}{h_j} \left( \hat{\gamma}(\bar{h}_j) - \gamma(\bar{h}_j)\right)^2
\end{equation}
It allows us to evaluate how well the different models fit the data and determine the best fit for the data by its minimum value.
The WSSE is more discriminant for the case of the $Z$ component than for the $V_Z$, where the exponential model seems to fit the data better. 

We also computed the Akaike information criterion (AIC) to compare the different possible models, but the AIC results were similar in all cases. This may be because all the models have the same number of parameters. 

Another model diagnostic is carried out by cross-validation, which lets us test the quality of the variogram fitting and compare among the different models. We took a $n_{f}=5$-fold partition of the data, i.e. $1/5$ of the data is taken as the modelling set $S_M$, to model the variogram and predict the locations of the remaining data, or validation set $S_V$. Then the validation measurements are compared to their predictions. We repeat this procedure a hundred times, where the modelling and validation sets are set randomly at each run. We estimate the goodness of the fit as the average of the sum of squared errors from the mean at each n-fold, 
\begin{equation}
    \label{eq:R2}
    R^2 = \frac{1}{n_{f}} \sum_{n}\left( 
    1 -\frac{\sum_{i\in S_{V_n}} (Z(r_i) - \hat{Z}(r_i))^2}{\sum_{i\in S_{V_n}} (Z(r_i) - \bar{Z}(r_i))^2} 
    \right)
\end{equation}
$R^2$ indicates how much Kriging prediction is a better predictor than the mean. We find it to be around the $99.9\%$ for all models in the $Z$ component. Similarly to the AIC criterion, this statistic is not discriminant for this variable. The results for $V_Z$ are nearly the same, being the Gaussian and the Spherical models slightly better. See Table \ref{tab:Table_AIC_WSSE}. 

We finally check the errors at each n-fold from the cross-validation analysis and how they are related with respect to the predictions $\hat{Z}(r_i)$ and the observed $Z(r_i)$, as shown in Table
\ref{tab:Table_AIC_WSSE}.
The cross-validation residuals $Z_{res} = Z(r_i) - \hat{Z}(r_i)$ are expected to be small for a correct variogram fitting. This is the case among all the models tested. 
The $z-$scores are the standardized residuals by the Kriging standard error \eqref{eq:sigma_krig}, $Z_{score} = (Z(r_i) - \hat{Z}(r_i)) / \sigma_{Z}(r_i)$, and they should have a mean close to 0 and variance values close 1. Again, we can check that all the variogram models are correct, and this statistic has a slight difference.   
We expect high correlation coefficients between the predictions and the observed data, named as $\hat{Z}(r_i) \sim Z(r_i)$.%
The higher the correlation, the better the model. 
While the correlation values between the residuals and predictions, ${Z}_{res} \sim \hat{Z}(r_i)$, are expected to be close to zero since the residuals should be random without any trend. 
The results seem to be tied in both cases, with the spherical model being slightly better for $Z$ and $V_Z$. 
Therefore, we chose the Exponential model for $Z$, based on a moderately better WSSE statistic, and the spherical model for the $V_Z$ variable. 

\section*{Data Availability}

The data underlying this article were accessed from \citet{Dias2021}, at \url{https://doi.org/10.1093/mnras/stab770}, and \citet{Alfaro22} at DOI: \href{https://iopscience.iop.org/article/10.3847/1538-4357/ac8b0c}{10.3847/1538-4357/ac8b0c}. The derived data generated in this research will be shared on reasonable request to the corresponding author.




\bibliographystyle{mnras}
\bibliography{Bibliography}{}



\bsp	
\label{lastpage}
\end{document}